# Boosting infrared energy transfer in 3D nanoporous gold antennas

D. Garoli,[a†] E. Calandrini,[a†] A. Bozzola,[a†] M. Ortolani,[b] S. Cattarin,[c] S. Barison,[c] A. Toma[a] and F. De Angelis*[a]

The applications of plasmonics to energy transfer from free-space radiation to molecules are currently limited to the visible region of the electromagnetic spectrum due to the intrinsic optical properties of bulk noble metals that support strong electromagnetic field confinement only close to their plasma frequency in the visible/ultraviolet range. In this work, we show that nanoporous gold can be exploited as plasmonic material for the mid-infrared region to obtain strong electromagnetic field confinement, co-localized with target molecules into the nanopores and resonant with their vibrational frequency. The effective optical response of the nanoporous metal enables the penetration of optical fields deep into the nanopores, where molecules can be loaded thus achieving a more efficient light-matter coupling if compared to bulk gold. In order to realize plasmonic resonators made of nanoporous gold, we develop a nanofabrication method based on polymeric templates for metal deposition and we obtain antenna arrays resonating at mid-infrared wavelengths selected by design. We then coat the antennas with a thin (3 nm) silica layer acting as target dielectric layer for optical energy transfer. We study the strength of the light-matter coupling at the vibrational absorption frequency of silica at 1250 cm$^{-1}$ through the analysis of the experimental Fano lineshape that is benchmarked against identical structures made of bulk gold. The boost of optical energy transfer from free-space mid-infrared radiation to molecular vibrations in nanoporous 3D nanoantenna arrays can open new application routes for plasmon-enhanced physical-chemical reactions.

## Introduction

Plasmonics has emerged because it has given the ability to manage light at the nanoscale by geometric manipulation of noble metal nanostructures. One of the key aspects of plasmonic nanostructures is the confinement of the optical energy into very small volumes – the so-called plasmonic hot spots – where the electric field is strongly enhanced compared to the electric field of the incoming free-space radiation, whose intensity cannot otherwise surpass its maximum value attained in a diffraction-limited focal spot. By using nano-antennas, metamaterials or other advanced electromagnetic concepts,[1–6] the hot spots can be engineered to concentrate optical fields down to volumes occupied by only few molecules, small nanoparticles or ultrathin films. When such nano-objects are placed into plasmonic hot spots, very efficient energy transfer can occur from the electromagnetic field to the molecules, if compared to the bare electromagnetic absorption cross-section of a molecule in a vacuum. Finding new field enhancement geometries that boost optical energy transfer can therefore trigger physical-chemical processes occurring beyond free energy thresholds that are otherwise difficult or even impossible to reach at the nanoscale[7-9]. Serious challenges to the efficiency of this approach, however, are posed by the extremely small modal volume and the inhomogeneous field distributions resulting from the typically low surface-to-volume density ratio of plasmonic hot spots in nanostructures. In fact, in many practical cases, the full technology exploitation of plasmons is hindered by the difficulties of filling with plasmonic hot-spots the entire volume occupied by the target nano-objects, such as catalytic reagents,[10] analytes for enhanced spectroscopy,[11] nanoparticles used as active gain medium for light emission,[12,13] or other. In other words, it can be stated that the co-localization of optical energy and target nano-objects in the same position is a major issue affecting the present performances of plasmonic devices.

Achieving efficient optical energy transfer may be even more difficult in the mid-infrared (mid-IR) region (wavelengths of approximately 2 to 12 microns), where the narrow-linewidth vibrational fingerprints of molecules would allow selective excitation of selected chemical species, but unfortunately the intrinsic advantages of plasmonics tend to vanish with increasing wavelength. This is due to a progressive and strong increase of the imaginary part of the noble metal permittivity $\varepsilon_2$, which is responsible for optical energy losses and plasmon damping, and a simultaneous increase of the absolute value of the real part of the permittivity $\varepsilon_1$, whose value close to unity is responsible for field confinement in the visible range. In fact, applications of plasmonics are currently limited to the visible/near-infrared region of wavelengths shorter than 1 micron and, although some recent progresses have been made using non-metallic conductors,[14-18] the research for alternative materials for mid-IR plasmonics is still an open issue. The options that have been explored up to now include conducting oxides, heavily doped semiconductors, phase-change amorphous materials and graphene. Conducting oxides include the Indium-tin oxide (ITO) and Al-doped ZnO (AZO), which are transparent to visible radiation, are easy to deposit with low-cost processes and feature widely tunable $\varepsilon_1$ and $\varepsilon_2$ in the entire IR range[14,15]. The alloy nature of ITO and AZO, however, makes their optical properties very sensitive to fabrication processes including material removal at extremely high spatial resolution (e.g. ion milling, reactive ion etching, etc.), which are needed to obtain plasmonic nanostructures. Heavily doped semiconductors with small effective mass like electron doped InAs [16] and Ge [17] are definitely more suitable for nanostructure fabrication and integration into microelectronic chips, but, with the doping levels obtained at the present state of the art of materials science, they can only function as plasmonic materials at far-IR wavelengths longer than 10 microns. Nanoantennas made of phase-change materials[18] have shown large tunability with temperature of the resonant wavelength in the mid-IR from 5 to 8 microns, but, for most chemical catalysis applications, temperature changes are precisely what one would like to induce with the optical energy transfer, rather than having to provide thermal energy to the plasmonic system through a heater. Graphene metasurfaces[19] have demonstrated high surface sensitivity, however at the expense of strongly reduced total sensing volume if compared to standard plasmonic metal meshes[20]. In summary, the problem of finding efficient materials and device concepts for mid-IR Plasmonics is not solved, and it will be even more urgent in the next future, because of the recent commercialization of quantum cascade lasers (QCLs)[21] that are now broadly tunable in the mid-IR range, enabling automated on-chip gas sensors based on vibrational spectroscopy. More importantly for the scope of this paper, the QCLs can make free-space mid-IR radiation beams with peak power of the order of 1 W available for optical energy transfer to specific molecular vibrational lines, due to their narrow emission linewidths of the order of few GHz or less[21].

In this work, we show that the use of porous loadable metal nanostructures can provide a straightforward method for the co-localization of molecules and mid-IR plasmonic hot-spots with extremely high surface-to-volume density ratio, insofar boosting the total integrated optical energy transfer from the mid-IR free-space radiation to the molecule vibrations. Due to the effective optical properties of the porous material in the mid-IR, which are found to be very different from those of bulk metal, the optical fields can penetrate deep into the metal nanostructure, while the molecules can fill the empty spaces created by the nanopore fabrication process. Efficient energy transfer between the free-space optical beam and the loaded molecules then occurs through mid-IR plasmon oscillations of the nanostructure. The plasmonic behavior in the mid-IR can be optimized at specific wavelengths through the design of nanostructure arrays and their subsequent characterization by IR spectroscopy, as we do here. In particular, here we show that, as in the case of bulk metals, it is possible to pattern the nanoporous metals deposited on-chip along the out-of-plane direction, thus achieving vertical nanoantenna arrays resonating at the desired wavelength. It is shown here that tuning the antenna resonance in the vicinity of the vibrational frequency of the molecules, together with increasing the interaction volume in the out-of-plane nanostructure configuration,[22] further strengthens the coupling between the plasmons and the molecules loaded into the metal nanopores.

## Results and discussion

Nanoporous gold (NPG) has been known for many years but it is still attracting strong interest due to the unique properties resulting from its high specific surface area (a roughness factor up to 20-25 has been estimated for a porous gold layer obtained under conditions similar to ours[23]). Nanoporous metals find applications in many fields from electrochemistry[24] to nanofluidics[25] and enhanced spectroscopy[26-31]. The preparation of NPG structures is based on physical-chemical procedures[32,33] where Ag is selectively leached from (Au,Ag) alloys. NPG thin films (Fig. 1 (a)) were prepared following the procedure described in Experimental Section. The optical response in the mid-IR was investigated by means of IR reflectance measurements (Fig. 1 (b)). A Drude-Lorentz model was fitted to the experimental data in order to derive the effective complex dielectric function of this specific NPG sample (Fig. 1 (c) and (d)).

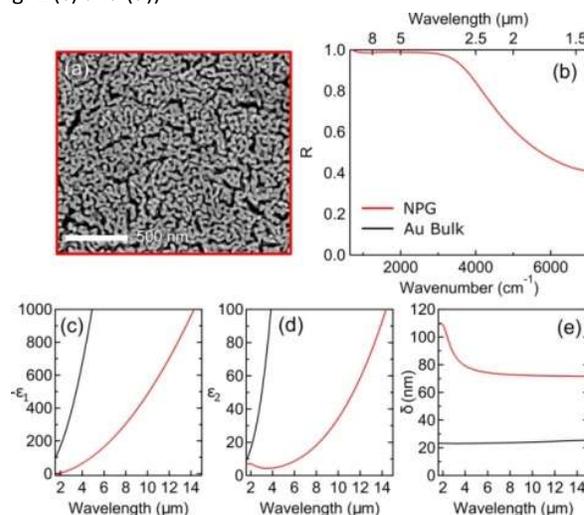

*Figure 1. Permittivity of nanoporous gold (a)* SEM image of an investigated NPG sample obtained with a de-alloying time of 3 hours. *(b)* Measured FTIR reflectance spectra at quasi-normal incidence. Real *(c)* and imaginary *(d)* part of the permittivity, and skin depth *(e)* estimated with the Drude-Lorentz model. The reference spectra of bulk gold are also reported (black lines).

It is well known[32] that, according to the specific porosity level, NPG can behave like an effective conductor with plasma frequency $\omega_p$ in the mid-IR range. The main hypothesis of the effective-medium approximation, that the structural feature size is much smaller than the wavelength, is acceptable in the mid-IR and near-IR range, however conventional effective medium theories, like the Maxwell-Garnet theory or the Bruggeman theory, fail to reproduce the NPG optical constants, because of the high degree of connectivity of the nanostructures and more importantly, because of their high shape variability.[32] Therefore, the obtained effective-medium values of the real (Figure 1 (c)) and imaginary (Figure 1 (d)) parts of the effective permittivity in the IR range have to be considered as phenomenological values[34], and the relation of the obtained Drude parameters (free carrier density and mass, carrier relaxation time, screening constant) to the geometric parameters of the material (pore size and distribution) and to the bulk gold dielectric function will be a subject of a forthcoming paper. The observed reduction of the absolute values of the dielectric permittivity of NPG if compared to that of bulk gold has important consequences. First, the reduced $\varepsilon_2$ enables lower ohmic losses. Second, the reduction of the imaginary part of the refractive index κ (which depends on both the real and the imaginary parts of the dielectric function) implies an increase of the skin depth $\delta=c/\omega\kappa$, which, in turn, enables the IR radiation to penetrate deep into the buried nanopores.

A quantitative estimate of the skin depth is plotted in Figure 1 (e). The skin depth, around 80 nm, is almost four times larger than the value of bulk gold (less than 25 nm in the IR)[34], hence the electromagnetic field penetrates into the buried nanopores where both the plasmonic hot spots and the nano-objects of interest are present, ultimately realizing the co-localization of optical energy and molecules.

In order to exploit the co-localization of optical energy and vibrational dipoles, we fabricate mid-IR plasmonic nanoantennas made of NPG. First of all, we simulated different square arrays of vertical nanoantennas made of both bulk gold and NPG. The height of the antennas is set by the present fabrication process to 2.3 µm, while the lattice pitch $P$ varies in the range 2.5-4.0 µm that generate resonances in the mid-IR as we now explain. A TM-polarized plane wave (magnetic field **H** orthogonal to the axis of the antennas) incident at 30° is used in the simulations as well as in the subsequent experiments. A full 3D electromagnetic calculation that takes into account the real geometry of the nanopores is currently a very hard challenge that would require a huge computational effort, therefore we treated the porous metal as a uniform material with dielectric function given by the effective phenomenological values experimentally determined in Fig. 1. The results for the field enhancement $|\mathbf{E}/\mathbf{E}_0|$ when the antennas are brought at resonance are illustrated in the vertical section of the field map shown in Figure 2 for $P$ = 3.5 µm. The deeper penetration of the electromagnetic field inside the volume of the vertical antenna can be clearly observed.

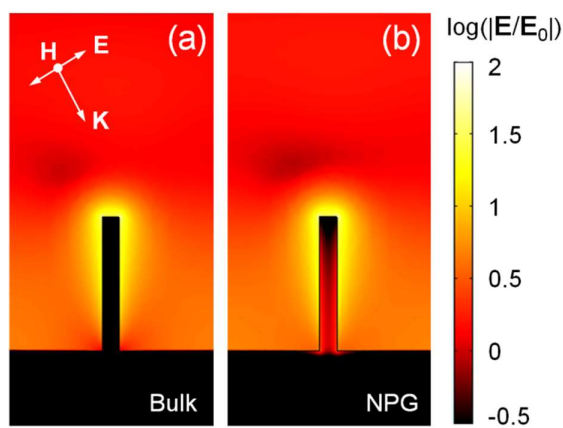

*Figure 2. Electromagnetic simulation of the electric field intensity at resonance in the elementary cell of a vertical nanoantenna array. The vertical antenna height is 2.3 µm and the array period is 3.5 µm. The logarithmic colour scale is the same for both panels. The resonance wavelengths are slightly offset (approximately 8 and 9 µm for bulk gold and NPG respectively) due to the dependence of the resonance wavelength on the different dielectric permittivity of bulk gold and NPG. The figure shows vertical sections through the antenna axes and the comparison between bulk gold and NPG demonstrates that the field penetrates more inside the NPG antenna.*

We now turn to the demonstration of enhanced transfer of optical energy to matter embedded in the NPG pores. First of all, we note that, due to the small size of the pores (tens of nanometers or less), patterning NPG into nanostructures for enhanced sensing/spectroscopy in the VIS spectral regions[6] can be difficult[27], because the size of the subwavelength features would be comparable to that of the pores. Instead, we now show that it is possible to fabricate micron-scale antennas for the mid-IR out of NPG by using properly designed 3D templates. Moreover, as said above, the approximation of a homogenous effective dielectric function is increasingly valid at longer wavelengths. The mid-IR nanoantenna fabrication process is as follows: we first fabricate polymeric templates of the vertical nanoantennas, namely polymeric pillars protruding from the substrate plane[22]; then we deposit the Ag-Au alloy over the templates; finally we leach the Ag from the alloy with a wet etch, thus realizing vertical out-of-plane antennas made of NPG and resonating in the mid-IR range. The fabrication process is described in details in the Experimental Section below. Examples of the fabricated nanoantennas are reported in Figure 3(a). The same structures, prepared for reference by depositing bulk gold on the same templates are shown in Fig. 3 (b).

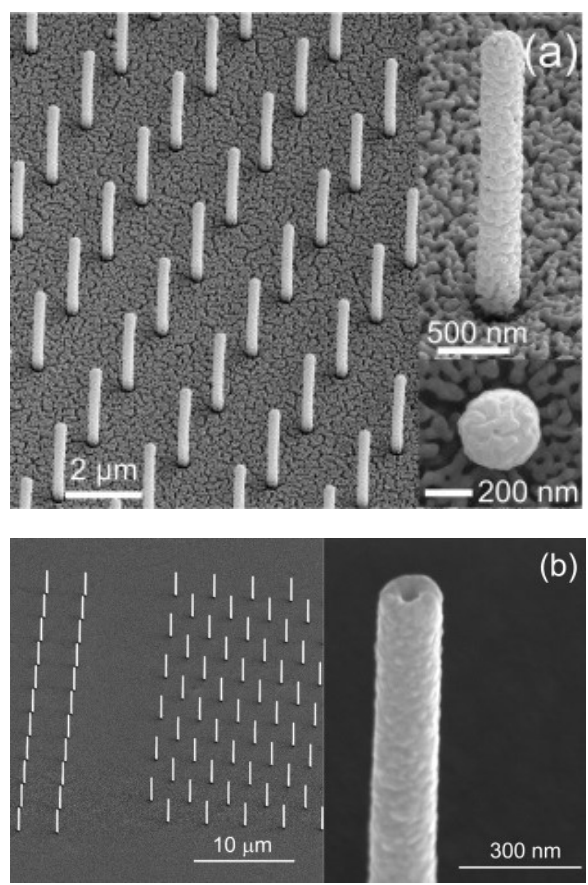

*Figure 3. SEM micrographs of arrays of antennas. (a) Characteristic array of out-of-plane vertical NPG antennas. In the top-right inset: side view. In the bottom-right inset: top view (notice the circular section). (b) Array of bulk gold antennas.*

To match the estimated skin depth value of NPG, the thickness of the porous metal is set to 80 nm, which gives a final geometrical aspect ratio of cylindrical antennas height/diameter ∼ 6. The porous structure of the metal is clearly visible in Fig. 3(a). We characterized the far-field optical response of the antenna arrays by means of Fourier-transform IR (FTIR) spectroscopy in a microscope with a 30°-incidence Cassegrain objective in reflection mode. At non-normal incidence, the radiation electric field vector has a non-zero component along the $TM_{10}$ mode, in which the electric field is mostly aligned along the out-of-plane direction, parallel to the main axis of the antenna hence exciting the fundamental antenna mode. The specular reflectance spectrum of the antenna array R(λ) was obtained by using a bulk gold-coated surface as reference. In Figure 4, the reflectance spectra of arrays with different pitch (height 2.3 µm, total radius 360 nm, array pitch varying in the range 2.5 – 4.0 µm) are reported together with the electromagnetic simulations of the reflected intensity, in which the illumination geometry of the FTIR experiment was also simulated. The reflectance dip displayed at the same wavelength by simulations in Figure 4 (a) and data in Figure 4 (b) confirms the validity of both our modelling and fabrication approaches. The interpretation of the reflectance dip is as follows: due to overlap of the square lattice resonance wavelength with the fundamental longitudinal mode of the antenna,[22] a maximum in the intensity scattered at non-specular angle S(λ) is expected. Since in the absence of transmission and for negligible absorption one has R(λ)∼1-S(λ), a dip is seen in the reflectance spectra of Fig. 4. The wavelength at which the dip appears redshifts proportionally to the increasing lattice period by more than 25%, because the fundamental mode of the vertical antenna has a bandwidth of more than 50% of the centre wavelength, therefore it is the array resonance that controls the exact spectral position of the dip and the linewidth (half width at half maximum is less than 10% of the centre wavelength).

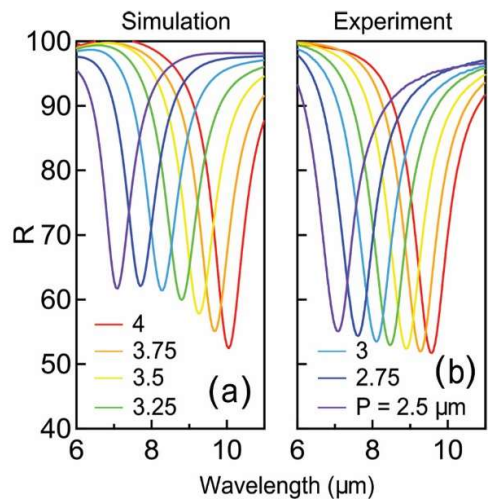

***Figure 4.*** *Optical response of arrays of antennas made of nanoporous gold. Simulated **(a)** and measured **(b)** reflectance of several arrays of NPG antennas with different lattice pitch P in the range 2.5 – 4.0 µm.*

The demonstration of the enhanced transfer of optical energy to vibrational energy is done by sensing the mid-IR vibrational modes of a thin layer of amorphous $SiO_2$ deposited on the antenna arrays by means of atomic layer deposition (ALD)[35,36]. With this technique, exactly the same thickness of analyte layer can be deposited on different antenna arrays made of either NPG or bulk gold. Moreover, the intrinsic homogeneity of the deposition rate of ALD ensures a uniform, conformal deposition of the analyte over the 3D surface formed by the porous structure[35,36]. As a demonstration of the uniformity and conformity of the ALD, we deposited the same analyte (about 10 nm of $SiO_2$, or 90 ALD cycles) on an NPG sample, and we imaged the cross-section by means of scanning electron microscopy (SEM). The image is reported in Figure 5 and it is possible to verify the uniformity of the $SiO_2$ layer across the entire 3D porous structure. For the sensing experiment, we choose to deposit 3 nm of $SiO_2$ on the nanoantenna arrays (30 ALD cycles).

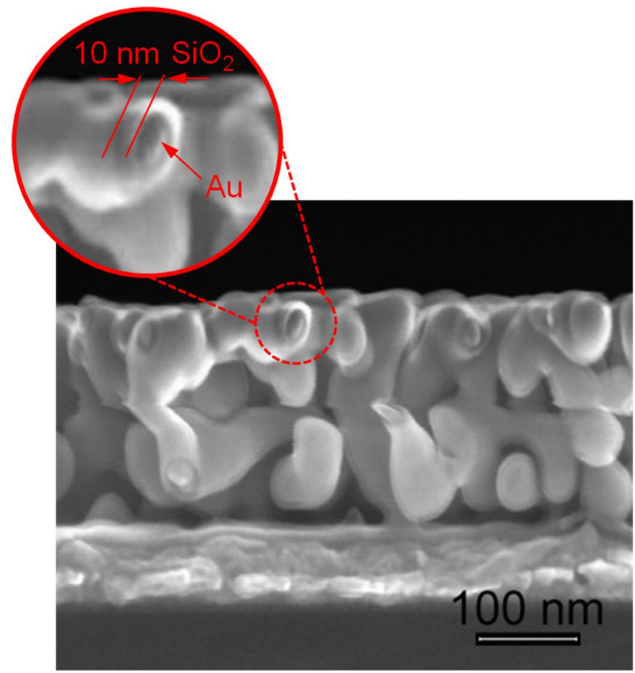

***Figure 5.*** *SEM micrograph of a NPG film cross section after ALD deposition of 10 nm of $SiO_2$.*

The sensing results are presented in Figure 6, where the FTIR reflectance microscopy spectra of the antenna arrays of Fig. 4 after the $SiO_2$ coating are reported in Figs. 6 (a) and 6 (b), respectively, as a function of the electromagnetic frequency ν=c/λ expressed in wavenumbers ω=1/λ. The antenna resonance could be tuned across the characteristic phonon energies of $SiO_2$ by changing the array pitch *P* as described in Fig. 4. $SiO_2$ is characterized by two distinct active phonons in the IR: a transverse optical (TO) phonon mode around 1090 $cm^{-1}$, and a longitudinal optical (LO) phonon mode at 1250 $cm^{-1}$.[37,38] According to the IR spectroscopy selection rules for a thin $SiO_2$ film on a flat surface, the TO phonon exhibits a strong thickness dependence, while the LO phonon is substantially thickness-independent and it can be excited only at non-normal incidence.[38] In a nanostructured amorphous sample, however, the selection rules are relaxed and both phonons are observed. The LO phonon produces a stronger signature than the TO phonon in the spectra of Fig. 6, so we shall focus on the LO vibration at $\omega_{LO}$=1250 cm$^{-1}$. The strength and the asymmetry of the spectral feature at $\omega_{LO}$ contains information on the light-matter interaction strength at the nanoscale, although, at the low photon flux of the FTIR blackbody source, the optical energy transfer is still negligible if compared to e.g. thermal energy at room temperature. We extract the light-matter interaction-strength parameters by performing a multi-peak fit to a broad Lorentzian background for the plasmonic resonance[39,40] and a narrow discrete vibrational transition line centred at $\omega_{LO}$ with an asymmetric Fano-Breit-Wigner profile[41,42]. In the spectral region around the antenna resonance, the $\omega$-dependent reflectance spectra are fitted to:

$$R(\omega) = R_0 - R_L(\omega) + R_F(\omega) \quad (1)$$

where $R_0$ is a constant baseline estimated in the range 95% – 100%, and $R_L(\omega)$, $R_F(\omega)$ are the Lorentz and Fano terms respectively, written as:

$$R_L(\omega) = \frac{A_L}{\pi} \cdot \frac{\gamma_L/2}{(\omega - \omega_{ant})^2 + (\gamma_L/2)^2} \quad (2)$$

$$R_F(\omega) = A_F \cdot \frac{[q + (\omega - \omega_{LO})/(\gamma_F/2)]^2}{1 + (\omega - \omega_{LO})^2/(\gamma_F/2)^2} \quad (3)$$

Here $A_L$ and $A_F$ denote the amplitudes of the Lorentz and Fano functions, while $\gamma_L$ and $\gamma_F$ denote the corresponding widths of the resonances. $\omega_{ant}$ is the antenna array resonance frequency, which is tuned in the range 900 – 1400 cm$^{-1}$ by changing $P$. The Fano-Breit-Wigner parameter $q$ gives the asymmetry of the line profile at $\omega_{LO}$, which is set by the light-matter interaction strength. An example of the results of the fitting procedure is reported in Fig. 6 (c) for the case of the NPG array with $P$=2.5 μm. In particular, the expression $1/q^2$ is a direct measure of the light-matter coupling coefficient[42] while $A_F$ measures the total energy transfer, therefore in Figs. 6 (d) and 6 (e) we plot $A_F$ and $1/q^2$ for all the bulk Au and NPG antenna arrays. It is clearly evident that the NPG arrays with $P$=3.0 μm and $P$=3.25 μm provide more than 4-fold enhancement in the optical energy transfer and in the strength of the light-matter interaction, respectively, if compared to other values of $P$. It may be interesting to notice that $A_F$ and $1/q^2$ are maximum when the plasmonic resonance frequency is slightly detuned from $\omega_{LO}$, e.g. the array with $P$=3.0 μm has the plasmonic resonance at 1150 cm$^{-1}$, -10% offset from $\omega_{LO}$, and $1/q^2$=46±10, while the fully-resonant array with $P$=2.75 μm and plasmonic resonance at $\omega_{LO}$ has $1/q^2$=3±1. This fact can be understood in the framework of more refined electromagnetic models that rigorously implement the role of both absorption and scattering enhancement in defining the asymmetric lineshape of molecules interacting with mid-IR nanoantennas[43,44]. The validity of our simplified analytical fitting formula, however, is confirmed by the values of the associated errors on $A_F$ and $1/q^2$, which are reported as vertical error bars in Figs. 6 (d) and 6 (e). The maximum values for the parameters $A_F$ and $1/q^2$ of the NPG arrays are always safely larger than the corresponding maximum values for the bulk gold arrays, which also show only 2-fold enhancement of optical energy transfer at $P$=3.25 μm and $P$=3.5 μm if compared to other values of $P$. We therefore conclude that the mid-IR radiation energy is most efficiently transferred to the vibrations of the SiO$_2$ layers deposited into the nanopores of the NPG structure with $P$=3.0 μm and $P$=3.25 μm.

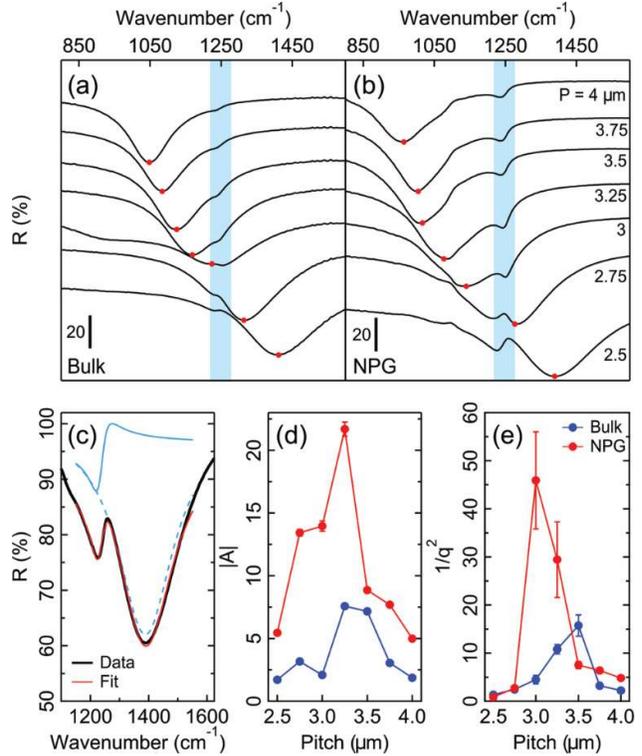

*Figure 6. Boost of the energy transfer in nanopatterned antennas arrays: NPG vs bulk gold.* *Reflectance spectra for several sets of arrays of bulk gold **(a)** and NPG **(b)** with different lattice periods (P). The positions of the Lorentz peak ($k_{ant}$) obtained by the multi-peak fit are marked with red dots, while the spectral window of the LO phonon in the SiO$_2$ is highlighted with a vertical blue shade. The spectra are shifted for clarity. **(c)** Example of the fitting procedure for the reflectance spectrum of NPG with P=2.5 μm. The measured curve is reported with a solid black line, the Lorentz curve with a dashed blue line, the Fano curve with a solid blue line, and the resulting sum with a solid red line. **(d)** The spectral weight of the Fano term for bulk gold (blue line and symbols) and NPG (red line and symbols) for different values of the array pitch P. **(e)** The light-matter interaction strength parameter $1/q^2$ for bulk gold (blue line and symbols) and NPG (red line and symbols) versus P. The errors associated to the fits are reported as vertical error bars in panels **(d)** and **(e)**.*

## Experimental

Thin films of NPG were prepared as follows: rectangular silicon substrates were thoroughly degreased in boiling acetone and dried. A 5 nm Ti layer was deposited on a silicon nitride membrane as an adhesion promoter. Subsequently an Au layer about 100 nm thick was deposited over Ti. Both these depositions were performed, using a Kenosistek® facility, in high vacuum (1 · $10^{-6}$ mbar) by means of electron beam (e-beam) evaporation at a deposition rate of 0.1 nm/s. Finally, a layer about 500 nm thick of $Ag_{75}Au_{25}$ alloy (at%) was deposited in a DC turbo sputter coater (Emitech K575X, Emitech Ltd., Ashford, Kent, UK), using a silver/gold alloy sputtering target $Ag_{62.3}/Au_{37.7}$ (wt.%), GoodFellow. The sputtering was performed at room temperature under Ar gas flow at a pressure of 7 · $10^{-3}$ mbar and a DC sputtering current of 25 mA. The composition of the alloy ($Ag_{75}Au_{25}$) was selected on the basis of literature reports discussed elsewhere[27-31], according to which the range 22–25 at.% Au is optimal. Etching baths were prepared by dilution of a concentrated $HNO_3$ solution to a final concentration of 33% (Sigma-Aldrich, ACS reagent 70%). The dealloying process was performed at room temperature for 3 hours. Each sample was then washed in distilled water, dried in a nitrogen stream and stored.

The method used to fabricate the samples relies on FIB-generated secondary-electron lithography in lithographic resists. S1813 optical resist (thickness of 2.3 µm) is spun at 3000 rpm on 200 nm thick $Si_3N_4$ membranes and soft-baked at 90°C for 4 minutes. On the back of the membrane a thin layer of gold (about 10 nm) is then deposited by means of sputtering. The membranes are then patterned from the backside using a Focused Ion Beam (Helios Nanolab600, FEI company), operated at 30 keV (current aperture: 40pA, dwell time: 500 µs, number of repetitions: 850 passes). Due to the high dose of low-energy secondary electrons induced by ion beam / sample interaction, a 40 nm thick layer of resist, surrounding the milled hole, becomes highly cross-linked and insoluble to most solvents. After patterning, the sample is developed in acetone, rinsed in isopropanol and dried under gentle N2 flow. A mild oxygen plasma (process time: 2 minutes, RF power: 100 Watt, gas flow: 25 sccm) reduces the polymeric skin down to 10 nm. A 80 nm thick layer of silver/gold alloy is deposited by sputtering the sample, tilted 60° with respect to the vertical and rotated, guaranteeing an isotropic coating on both the sidewalls and the base. The deposition follows the same parameters reported above for the preparation of the NPG thin-films. Finally, in order to prepare the sample with the desired nanoporous structure, the nanoantennas were etched in diluted $HNO_3$ for 3 hours at room temperature.

The optical response in the mid-IR is investigated by means of FT-IR reflectance microscopy measurements by using a Nicolet™ iS™ 50 FT-IR Spectrometer coupled to the Nicolet Continuµm Infrared Microscope by Thermo Scientific™.

Electromagneticx simulations are performed by means of the Comsol Multiphysics® software. Periodic boundary conditions are imposed at the lateral facets of the unit cell, and perfectly matched layers are assumed at the top and bottom boundaries of the simulation box.

## Conclusions

In summary, we have shown that nanoporous gold can provide a longer effective skin depth with respect to bulk gold, which can be beneficial for optical energy transfer to molecule vibrations in the mid-infrared. In particular, the longer penetration depth enables the co-localization into the pores of the vibrating molecular dipoles and the optical energy, whose transfer can be boosted by designing suitable plasmonic resonators with high hot-spot density. Here we designed, fabricated and spectroscopically characterized vertical nanoantenna arrays made of nanoporous gold, which were benchmarked to identical structures made of bulk gold fabricated by using the same polymeric templates. The nanoantenna arrays displayed a plasmonic response tunable by the array pitch and an enhanced light-matter coupling to a 3 nm thick conformally deposited silicon dioxide layer. The observed efficient energy transfer from the free-space mid-infrared beam to the vibrations is explained by the co-localization of the nanometric layer and the plasmonic hot spots inside the pores, thus opening a new route towards plasmon-enhanced physical-chemical effects featuring high molecular specificity typical of the mid-infrared range.

## Acknowledgements

The research leading to these results has received funding from the European Research Council under the European Union's Seventh Framework Program (FP/2007-2013) / ERC Grant Agreement n. [616213], CoG: Neuro-Plasmonics and under the Horizon 2020 Program, FET-Open: PROSEQO, Grant Agreement n. [687089].

## Notes and references